\documentclass[aip, jcp,  amsmath,amssymb]{revtex4-1}
\usepackage{bm}
\usepackage{float}
\usepackage{graphicx}
\usepackage{natbib}
\usepackage{color}
\begin{document}
\title{Single-molecule Electronics: Cooling Individual Vibrational Modes by the Tunneling Current}
\author{Jacob Lykkebo}
\affiliation{Nano-Science Center and Department of Chemistry, University of Copenhagen, Universitetsparken 5, 2100 Copenhagen \O, Denmark}
\author{Giuseppe Romano}
\affiliation{Department of Mechanical Engineering, Massachusetts Institute of Technology, 77 Massachusetts Avenue, Cambridge, MA 02139}
\author{Alessio Gagliardi}
\affiliation{Technische Universit\"at M\"unchen, Electrical Engineering and Information Tech., Arcisstr. 21, 80333 M\"unchen, Germany}
\author{Alessandro Pecchia}
\affiliation{Consiglio Nazionale delle Ricerche, ISMN, Via Salaria km 29.6, 00017 Monterotondo (Rome), Italy}
\author{Gemma C. Solomon}
\email{gsolomon@nano.ku.dk}
\affiliation{Nano-Science Center and Department of Chemistry, University of Copenhagen, Universitetsparken 5, 2100 Copenhagen \O, Denmark}
\date{\today}
\begin{abstract}
Electronic devices composed of single molecules constitute the ultimate limit in the continued downscaling of electronic components. A key challenge for single-molecule electronics is to control the temperature of these junctions. Controlling heating and cooling effects in individual vibrational modes, can in principle, be utilized to increase stability of single-molecule junctions under bias, to pump energy into particular vibrational modes to perform current-induced reactions or to increase the resolution in inelastic electron tunneling spectroscopy by controlling the life-times of phonons in a molecule by suppressing absorption and external dissipation processes.
\\
Under bias the current and the molecule exchange energy, which typically results in heating of the molecule. However, the opposite process is also possible, where energy is extracted from the molecule by the tunneling current. Designing a molecular "heat sink" where a particular vibrational mode funnels heat out of the molecule and into the leads would be very desirable. It is even possible to imagine how the vibrational energy of the other vibrational modes could be funneled into the "cooling mode", given the right molecular design. 
\\
Previous efforts to understand heating and cooling mechanisms in single molecule junctions, have primarily been concerned with small models, where it is unclear which molecular systems they correspond to. In this paper, our focus is on suppressing heating and obtaining current-induced cooling in certain vibrational modes. Strategies for cooling vibrational modes in single-molecule junctions are presented, together with atomistic calculations based on those strategies. Cooling and reduced heating are observed for two different cooling schemes in calculations of atomistic single-molecule junctions. 
\end{abstract}
\maketitle
\begin{figure}[H]
\centering
\includegraphics[width=0.75\textwidth]{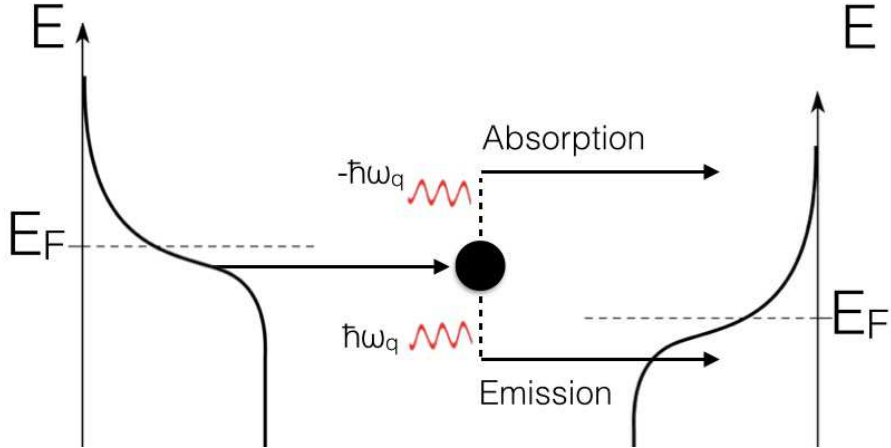}
\caption{Examples of inelastic tunneling \textit{via} absorption or emission of a phonon. The molecule is represented by the black circle in the middle and the semi-infinite leads by the Fermi distributions. The finite temperatures of the leads are implied by the energy dependence of the Fermi functions. }
\label{heating_cooling}
\end{figure}

%
%
Since the first measurement of the conductance of a single molecule\cite{reed1997}, the field of molecular electronics has attracted enormous attention. A crucial feature of any future device is the stability. In this paper, our focus is on the non-equilibrium effective temperature of a single-molecule junction induced by the tunneling current, and strategies to design molecules where the effective temperature of particular vibrational modes can be cooled below the ambient temperature. Although beyond the scope of this paper, it can be imagined that if a particular vibrational mode was tuned such that the absorption of the phonon energy by the current was favored over the emission into the mode, that this mode could act as a \textit{heat sink} for the vibrational energy put into the \textit{other} modes. While the idea of current-induced cooling is not new \cite{galperin2009, mceniry2009, arrachea2012}, the purpose of this article is to go beyond simple model systems to realizable molecules. Heating of single molecules has been inferred indirectly \cite{huang2006, huang2007, schulze2008}, and directly from the combination of surface-enhanced Raman spectroscopy and transport measurements\cite{ioffe2008,ward2011}, which paves the way for a detailed understanding of non-equilibrium heating and cooling.
\\
We proceed by first considering simple model systems in which the necessary molecular properties are established. In the second part of the paper, we go beyond the simple model calculations to an atomistic description. Molecules that meet the requirements are designed, and a slight cooling effect is observed. While the cooling effect presented here is small, we believe that the design strategies outlined in this paper could aid in the rational design of single-molecule devices with increased stability.
%
%
%
We obtain the power pumped into a molecule in a junction under bias by a semi-classical rate-equation balancing the net rate of phonons going into mode $q$\cite{paulsson2005, pecchia2007, romano2010}.
\begin{equation}\label{steadystate}
\frac{dN_q}{dt} = (N_q + 1)\cdot E_q - N_q\cdot A_q \pm J_q(N_q - n_q(T_0)).
\end{equation} 
Here, $N_q$ is the non-equilibrium phonon population of the vibrational mode, $q$, from which the effective local "temperature" of mode $q$ can be extracted using the assumption
\begin{equation}
N_q = n_q(T_q), 
\end{equation}
which implies that an effective temperature can be extracted from the non-equilibrium phonon population in mode, $q$ assuming a Bose-Einstein distribution. Emission, $E_q$, of a phonon can occur when the molecule is in either the vibrational ground state or the vibrationally excited state, whereas absorption, $A_q$, of a phonon can only happen \textit{via} an excited state. This asymmetry is evident from the pre-factors. The heating process is thereby in general favored over the cooling process. The rates for $A_q$ and $E_q$ are calculated using the non-equilibrium Greens function method. $J_q(N_q - n_q(T_0))$ describes the exchange of energy between the leads and the molecule and is controlled here by a single pre-set parameter, $J_q$, which describes dissipation to and from the leads \cite{pecchia2007,romano2010}. The temperature of the leads is assumed to be constant. The details of our method has been described in detail elsewhere \cite{pecchia2007, gagliardi2008, romano2010}.
\\
At steady-state ($\frac{dN_q}{dt}=0$) the effective temperature of a vibrational mode $q$ is given by
\begin{equation}
N_q = \frac{n_q(T_0)J_q + E_q}{J_q + A_q - E_q}
\end{equation}
This temperature cannot of course have all the thermodynamic properties of temperature in thermodynamic equilibrium, however it is possible to show that they share some of the conventional temperature properties, \textit{i.e.} it is always positive and an increasing function of the energy. It represents the temperature at which the system (the molecule) would relax to if suddenly disconnected from the contacts. This parameter is sufficient for our discussion here concerning cooling and heating of the molecule in terms of increase/decrease of energy in molecular vibrations \cite{galperin2011, narayanan2012, meair2014, gagliardi2015}. An extensive debate about the concept of temperature in non-equilibrium, can be found elsewhere \cite{casas-vazquez2003, dubi2011}.
\\
The current-induced heating and cooling processes can be described to first order by a number of terms that involve tunneling \textit{via} a scattering channel from one of the leads into the molecule, followed by an absorption or emission of a phonon, before tunneling \textit{via} another scattering channel out to the leads. Heating and cooling can occur from tunneling of electrons or holes as well as from virtual processes. Two of these processes are illustrated in figure \ref{heating_cooling}. Higher order processes are also possible, but only first order processes are expected to \cite{hipps2012} and found to be relevant in this study. 
%
\subsection{Results and Discussion}
\begin{figure}[H]
\centering
\includegraphics[width=1.0\textwidth]{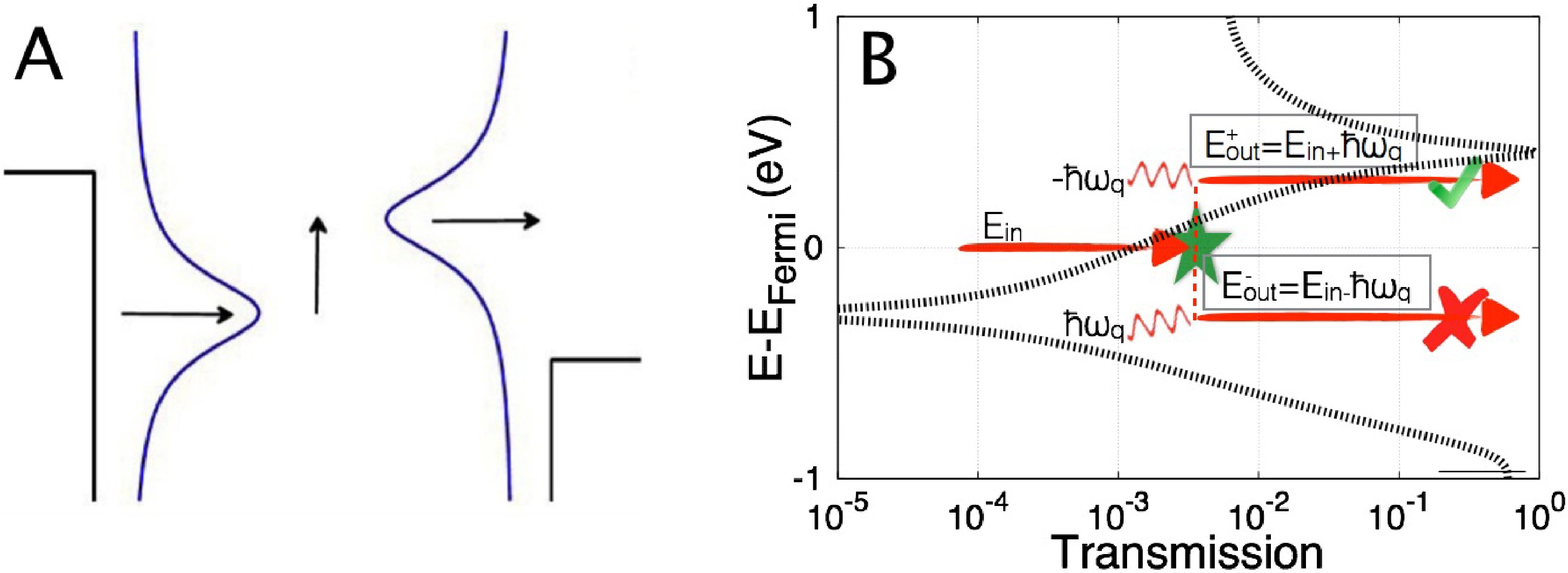}
\caption{A: Scheme A, Two weakly coupled molecular subunits with slightly different energies are bridged by a vibrational mode. B: Scheme B. A phenomenological visualization of absorption and emission, overlaid the elastic transmission. Destructive interference blocks phonon emission. The red arrows illustrates phenomenologically the inelastic tunneling current across the molecule.}
\label{schemes}
\end{figure}

\begin{figure}[H]
\centering
\includegraphics[width=1.0\textwidth]{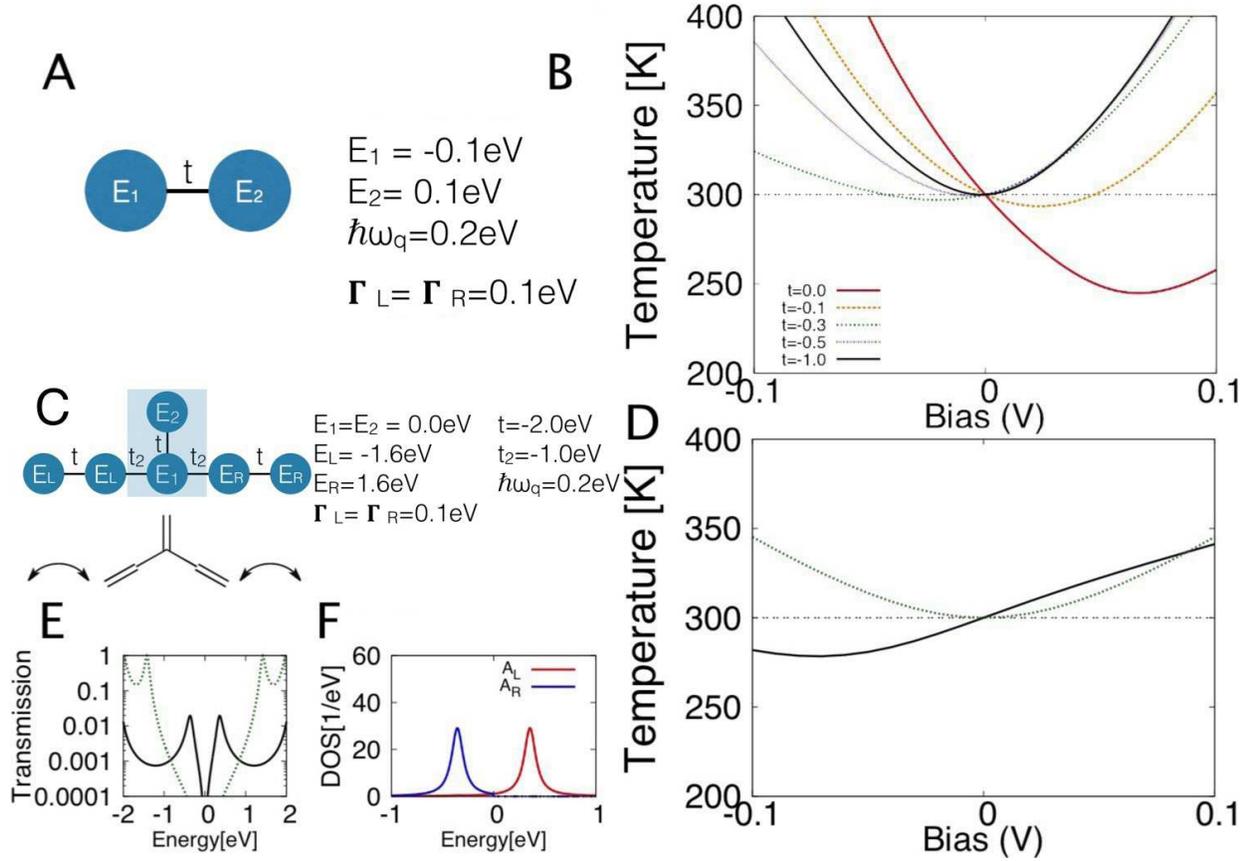}
\caption{Model results for scheme A. A: The connectivity of the simple two site model as well as the site energies and couplings used. B: Temperature vs. bias for the model. C: The connectivity of the six site model as well as the site energies and couplings used. The electron-phonon coupling matrix acts on the atoms in the shaded area. D: Temperature vs. bias for the model. The dotted line is for the case where all site energies are set to zero. E: Transmission of the six site model. The dotted line is for the case where all site energies are set to zero. F: The energy dependence of incoming and outgoing scattering channels.}
\label{model_results}
\end{figure}

\begin{figure}[H]
\centering
\includegraphics[width=1.0\textwidth]{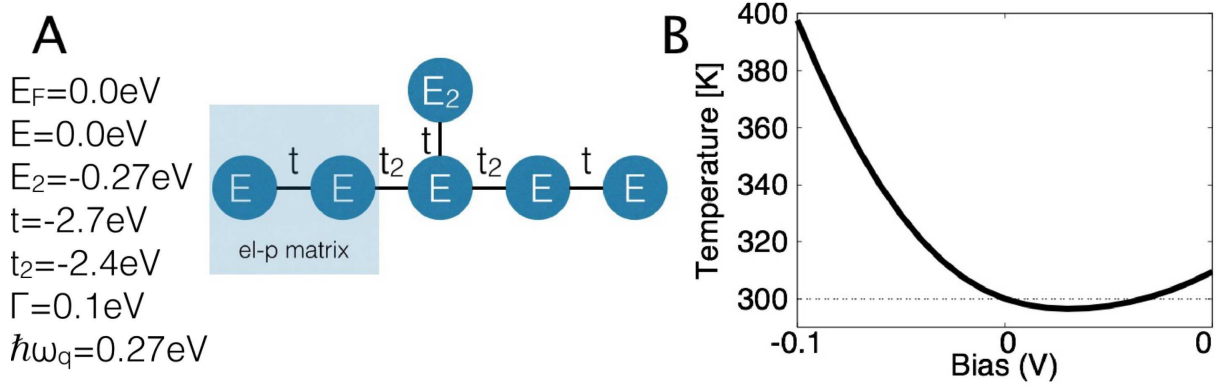}
\caption{Model results for scheme B. A: The connectivity of the model as well as the site energies and couplings used. B: Temperature vs. bias for the model.}
\label{model_resultsB}
\end{figure}

\textbf{Basic Idea}. In order to obtain current-induced cooling, we must suppress emission and enhance absorption, and here two basic strategies are presented. In scheme A, the idea is to have an incoming and an outgoing channel that are coupled by a vibrational mode but poorly coupled electronically. This is achieved by tuning the energies of the channels such that absorption is favored over emission. At low bias, the emissive backscattering process should be suppressed by the Fermi population in the leads \cite{romano2010}. The idea is outlined in figure \ref{schemes}A.
\\
In scheme B, the idea is to exploit the energy dependence of the transmission in a system with a destructive interference feature near the Fermi energy. The setup consists of a vibrational mode, which interacts strongly with the current followed by a molecular motif which causes destructive interference. When the incoming current interacts with the vibrational mode, the energy of the outgoing current will be shifted up (down) by the absorption (emission) event. The tunneling event after the phonon interaction will therefore leave the molecule at two different energies. By positioning the destructive interference feature in energy such that the interference dip occurs at an energy equal to the Fermi energy \textit{minus} the phonon energy the emission process can be suppressed. In this design we essentially assume that the inelastic transmission after emission is similar to the elastic transmission only shifted by phonon energy. We shall return to this point below. The idea is outlined in figure \ref{schemes}B.
\textbf{Tight-binding Calculations}. Scheme A can most easily be realized by having two sites, with slightly different site energies connected \textit{via} a vibrational mode. This model has been studied in relation to current-induced cooling previously \cite{lu2011, arrachea2014, simine2012}. The details of the model and the results are shown in figure \ref{model_results}A-B. The vibrational mode, $q$ with energy $\hbar\omega_q$ corresponds to a coupling between the two sites and is a simplified version of the electron-phonon coupling matrix for the C=C double bond stretch mode obtained from an atomistic calculation \cite{lykkebo2013}. The electron-phonon coupling matrix, $\bm{\alpha}$, has the form
\begin{equation}
\bm{\alpha} = 
\begin{pmatrix}
0 & 0.01 \\
0.01 & 0 
\end{pmatrix},
\end{equation}
and is applied to the marked atoms in figures \ref{model_results}(A and C) and \ref{model_resultsB}A. Consider initially the $t$=0 situation in figure \ref{model_results}B, where the left and right channels are completely decoupled. In this case, the left (right) scatting channels are simply described by a broadening due to the coupling to the leads, $\Gamma_{L(R)}$, and a maxima given by the site energy of $E_L(E_R)$. The site energies are set such that $E_L + \hbar\omega_q = E_R$, and absorption is thereby favored over emission. It can be seen in figure \ref{model_results}B, that the phonon mode is cooled by the current in the positive bias direction, and as expected, the phonon mode is heated in the negative bias direction. We emphasize that this model system is very delicate, and slight changes to the system will remove the asymmetry in the molecule and hence the cooling effect. Changing the coupling element between the two sites, $t$, to a finite value, rapidly causes the scattering channels to become aligned, and thereby removes the cooling effect. This is shown for a range of coupling constants in figure \ref{model_results}B, the cooling effect rapidly goes away and the heating process dominates.
\\
While the first model system serves to illustrate that, under certain conditions, it is possible to make a system that cools down, it is clear that the system is quite artificial. In a two site H\"uckel Hamiltonian (with identical site energies) the HOMO-LUMO gap scales with the coupling strength, and as the cooling only works when the coupling is set to zero, it is clear that this model does not represent a real molecule.
\\
The question is then how to decouple the two scattering channels without resorting to unrealistic parameters? One way, is to employ destructive quantum interference. This is achieved by setting up a six site H\"uckel Hamiltonian, to emulate a cross-conjugated molecule. Expanding to six sites, and not just four, is necessary to avoid the system being a bi-radical. Again, the same single vibrational mode is used, but this time it mimics the stretch mode of the central two sites in figure \ref{model_results}C. At the Fermi energy, the incoming and outgoing scattering channels are completely decoupled and the transmission goes to zero. Ideally, the in- and outgoing scattering channels should have narrow peaks and their maxima should be separated by the energy of the phonon frequency.
\\
The molecular motif and the corresponding H\"uckel model are shown in figure \ref{model_results}C. We use different coupling elements in order to simulate double and single bonds. The bias-dependent temperature of the system is shown in figure \ref{model_results}D. In this case, it is significantly more difficult to find realistic parameters to obtain cooling, because it is more difficult to align the scattering channels. The transmission through the model is shown in figure \ref{model_results}E. For reference, the dotted line shows the transmission when all site energies are equal (0eV). When the energies of sites $\text{E}_L$ and $\text{E}_R$ differ, the HOMO-LUMO gap narrows (figure \ref{model_results}E, full line) and the peaks in the scattering channels are identical to the peaks in the transmission function. Thus, for the scattering channels to have any significant difference on the scale of the phonon frequency, it is necessary to have a rather large energy difference between $\text{E}_L$ and $\text{E}_R$. For the symmetric molecule, with $\text{E}_L$ equal to $\text{E}_R$, only heating is observed, whereas when $\text{E}_L = -\text{E}_R = -1.6eV$ the cooling effect from the simpler model is recovered. As a slight curiosity, the cooling is found to be in the opposite bias direction compared with the previous model. This is due to the energy dependence of the scattering channels, figure \ref{model_results}F, where $A_L$ ($A_R$) has a peak slightly above (below) the Fermi level.
\\
This second version of scheme A illustrates the technical difficulties in obtaining current-induced cooling. However, this model also provides the ingredients that is needed in order to approach the regimes where a cooling effect might be possible. Particularly, it is necessary to introduce an asymmetry into the molecule such that the scattering channels favors absorption over emission, and crucially the channels must be decoupled, which can be obtained with either interference effects or by twisted geometries, where orbital overlap is reduced, \textit{e.g.} 2-2'-dimethylbiphenyl type molecules \cite{ballmann2012, markussen2014}.
Scheme B: Cooling can also be obtained by exploiting destructive quantum interference directly. This is done by using quantum interference to selectively suppress the heating processes \textit{via} tunneling, by carefully aligning the energy of the incoming electron, the energy of the destructive interference feature with the phonon frequency it is possible to suppress the inelastic transport processes involving phonon emission. The idea is illustrated in figure \ref{model_resultsB}A, and requires the following condition
\begin{equation}
E_{in} - \hbar\omega_q = E_{qi} = E_{out}^{-}, 
\end{equation}
where $E_{in}$ is the energy of the incoming energy, at small bias essentially the Fermi energy. $E_{out}^{-}$ is the energy of the outgoing tunneling electrons after emission. $E_{qi}$ is the energy of the destructive interference feature and $\hbar\omega_{q}$ is the energy of the phonon frequency. It should be noted that coherence in general is lost in the absorption or emission process, and that one cannot be guaranteed to find the interference feature at the expected energy simply by studying the elastic transmission. However, for the model systems considered here, we find that the outgoing inelastic transmission is similar to the elastic transmission only shifted in energy, in agreement with the simplistic schematic shown in figure \ref{schemes}B.
\\
In this scheme, the site energies and couplings strengths can be kept at standard values, with the exception of the single site-energy, $E_2$, which determines the position of the interference feature. The parameters and connectivity of the model are shown in figure \ref{model_resultsB}A, and the temperature as a function of bias is shown in figure \ref{model_resultsB}B. Here, we consider only a single localized vibrational mode which is localized on the two sites in the shaded box in figure \ref{model_resultsB}A and this mode corresponds to a triple bond stretch mode. It should be noted that for a symmetric molecule, a similar vibrational mode is present on the two outermost sites on the right hand side of the molecule and with the same energy, thereby making it impossible to distinguish the effective temperature of each mode (This is illustrated in the supplementary information (SI), figure 1). However, in the tight-binding calculations presented in \ref{model_resultsB}, we choose to only consider a single vibrational mode, and in the atomistic calculations presented below, we ensure to construct our molecule so that only a single triple bond stretch is present. 
\\
It can be seen that a cooling effect is obtained in the positive bias direction, whereas a significant heating effect is observed in the negative bias direction. Here, we choose a phonon energy of 0.27eV which approximately equals the phonon energy of a carbon-carbon triple-bond stretch mode.
\begin{figure}[H]
\centering
\includegraphics[width=1.0\textwidth]{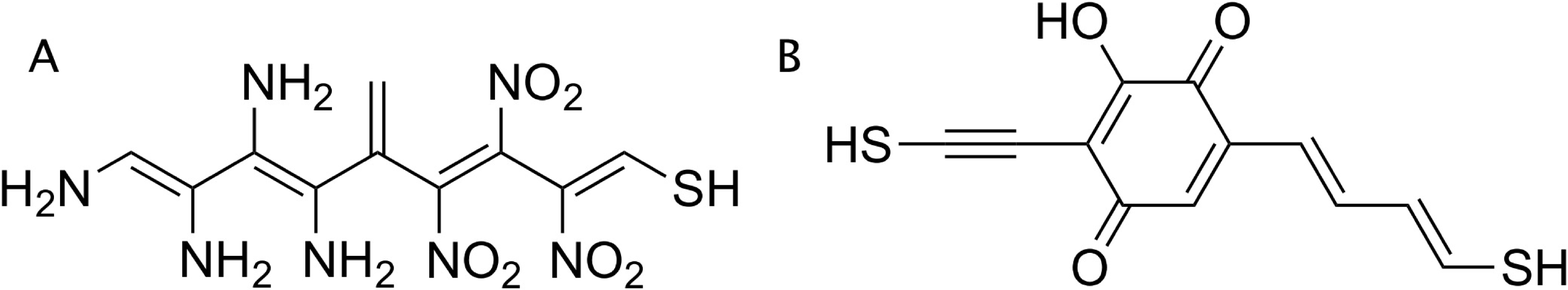}
\caption{A molecular realization of scheme A (A) and scheme B (B). Hereby named molA and molB, respectively.}
\label{molecules}
\end{figure}
\textbf{Atomistic Calculations}. To design a molecule that mimics scheme A, we must choose a molecule where the incoming and outgoing channels are decoupled, and where the channel energies differ significantly. In order to decouple the scattering channels, the central part of the molecule is composed of a cross-conjugated unit, which causes destructive quantum interference close to the Fermi level \cite{solomon2008}. The central unit is flanked by two linking groups that serve two purposes: To suppress the $\sigma$-transmission and tune the incoming and outgoing scattering channels. Molecules with a conjugated carbon backbone have transmission channels through both the $\pi$- and the $\sigma$-orbitals \cite{solomon2008, lykkebo2013}. Elongating the molecule reduces the relative contribution from the $\sigma$-orbitals, since they decay faster with length, than those from the $\pi$-orbitals (e.g. $\beta \sim$ 0.77$\text{\AA}^{-1}$ \: in $\sigma$-orbital dominated alkane-chains\cite{venkataraman2006b} compared with $\beta \sim$ 0.37$\text{\AA}^{-1}$ for the $\pi$-orbital dominated OPE series \cite{kaliginedi2012,valkenier2014}). The linker groups also provide positions where substituents can be placed to alter the energies of the conducting orbitals of the backbone, analogous to the site-energies of the H\"uckel-calculation. We choose to employ nitro ($-\text{NO}_2$) and amino ($\text{NH}_2$) groups, as these groups are on the opposite ends of the Hammett scale \cite{hammett1937}. Due to the many nitro and amino groups the molecule is highly polar, probably zwitterionic and possibly highly unstable. The molecule serves only to illustrate how a rational design following the guidelines can be tailored to suppress heating and favor cooling processes. 
\\
In scheme B, an interference feature is required to be separated from the Fermi energy, by exactly the energy of the phonon frequency such that emission is blocked. To obtain this relation, we need a molecule that exhibits destructive quantum interference, and some suitable linking groups to position the interference feature relative to the Fermi energy. In this study we use 1-4-benzoquinone as the central unit which is predicted to have destructive quantum interference near the Fermi energy \cite{markussen2011,lykkebo2014}. To fine-tune the interference feature into position we use a single alcohol substituent to position the interference feature correctly relative to the Fermi level within the approximation of our method \cite{andrews2008,lykkebo2014}. Our focus here is exclusively on the effective temperature of the single triple bond stretch mode with a phonon energy of $\sim 0.27eV$. Since the carbon-carbon triple bond stretch mode is relatively isolated compared with the other vibrational modes in an organic molecule, this mode might serve as a useful test case for observing current-induced cooling or suppressed heating. The molecules are shown in figure \ref{molecules} and are henceforth named molA and molB.
\\
The choices made for the molecules only serve as examples, and there are many choices to obtain the two requirements. Particularly, meta-coupled benzenes or quinone-type molecules, are molecular motifs that are know to have quantum interference effects \cite{fracasso2011,guedon2012, arroyo2013, arroyo2013b}. 
\\
We consider only a single carbon-carbon double bond stretch mode for the molecule in scheme A and the single carbon-carbon triple bond stretch mode for the molecule in scheme B. These modes mimics the phonon modes of the tight-binding calculations in the previous section. Due to the large size of the molecular systems, we restrict the calculations to only include first order processes. However, we find for smaller systems that these results correspond essentially to the results obtained while including higher order processes (See the SI, figure 2).
\\
%
The results for the two molecules are shown in figure \ref{abinitio_results}. The results are shown for three initial temperatures, and both a positive and a negative bias voltage is applied. Additionally, two control molecules are employed that are missing essential elements for the cooling effect. In scheme A, the central feature of the cooling design is the substituents, which tune the energies of the scattering channels. The control molecule, is similar to molA, but stripped of the substituents, aside the amine group that binds to the lead. In scheme B, the central feature of the cooling design is the destructive interference feature, so the control molecule is the corresponding hydroquinone molecule. This approach is know to remove the destructive interference effect \cite{markussen2010b, koole2015}. 
\\
For both molA and molB it is found that heating is significantly reduced at an initial temperature of 300K, and for molB a slight cooling effect is even observed at small bias. As the bias increases, the temperature rises, but far less in the positive bias direction compared with the negative bias direction. At small bias below the voltage threshold defined by the phonon frequency and low temperature, emission by tunneling electrons is suppressed due to the electronic occupation in the right lead\cite{romano2010}. It could therefore be hypothesized, that a cooling of the molecule could be observed at low bias and low temperature. However, since the net absorption rate is proportional to the phonon population, which in turn depends on the temperature of the system, the net absorption rate is very low, and hence emission still dominates (See the SI, figure 3). 
\\
For both systems it is observed that a robust and clear asymmetry in the biased induced heating is observed, which is a clear indication that it is possible to control electron-phonon interactions by rationally designing a molecule. Even when normalizing with respect to the current, the phonon mode is still colder in the positive bias direction compared with the negative. 
\\
For higher initial temperatures, the cooling effect is much more pronounced. This can be understood as the prefactor $N_q$ changes a lot more compared with the prefactor $(N_q+1)$, and thereby does the cooling effect become much more pronounced. Initially, the temperature rises monotonically with increasing bias, but at higher bias, the details of the transmission function begins to affect the results \cite{romano2010}.
\begin{figure}[H]
\centering
\includegraphics[width=1.0\textwidth]{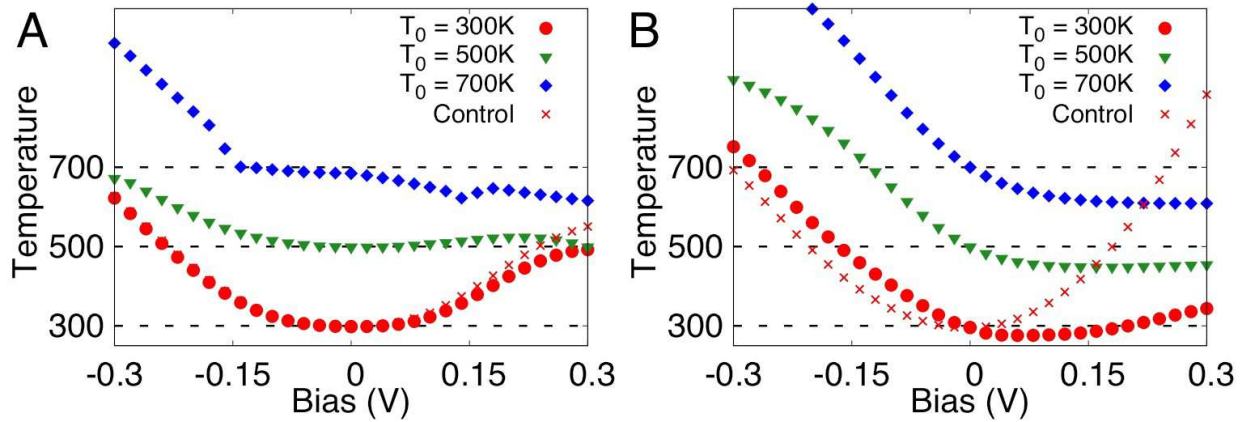}
\caption{Temperature vs. Bias for molA (A) and molB (B), respectively. The dashed horisontal lines indicate the three initial temperatures. }
\label{abinitio_results}
\end{figure}
\subsection{Conclusions and Outlook}
In this paper, we have discussed design strategies in order to enhance the inelastic absorption processes compared with the emission process leading to current-induced cooling. This is a desirable feature to control in molecular electronics devices. In simple tight-binding models, we show which design features that are necessary to cool a molecule by the current in two different ways. In the first scheme, the incoming and outgoing scattering channels must have different energy dependences such that the absorption of a phonon in a molecular vibrational mode is favored over emission of a phonon into the mode. In the second scheme, it is found that emission can be suppressed by directly utilizing destructive quantum interference. Previously, it has not been clear how to go from model calculation to molecule, but by following these strategies, it is more straight-forward to rationally design molecules where cooling or suppressed heating might be observable. Using these design strategies, suppressed heating and even a slight cooling effect was observed in atomistic calculations. The strategies outlined in this paper may therefore be employed to increase the stability of molecular junctions, thereby offering a possibility to increase the bias window in which they can operate.
\\
On the other hand, the design ideas outlined in this paper might also be utilized to design molecule where particular modes are selectively pumped with energy by the current. This can be done by reversing the bias polarity or by designing molecules with heating in mind instead of cooling. By doing this, it might also be possible to use the current to switch between different molecular structures, by either a geometrical rearrangement or by current-induced reactions. 
\subsection{Methods}
For the atomistic calculations, the electronic structures and electron-phonon couplings were obtained using the density functional tight-binding method \cite{porezag1995, elstner1998}, and all electron-transport calculations were performed using the non-equilibrium Greens function method \cite{pecchia2004}. The code utilizes a minimal basis set, which allows for reasonably fast results. The method however, has been found to yield reasonably accurate results when comparing to experiments of inelastic electron tunneling spectra \cite{solomon2006}. 
\\
In order to setup the atomistic calculations, some reasonable input parameters must be used. Particularly, the decay rate of the phonons into the leads, $J_q$. This parameter requires some special attention. The decay rates of the phonon modes directly into the leads can be calculated from first principles \cite{pecchia2007, romano2007, gagliardi2008}. However, the phonon density of states of the bulk gold has a cut-off at low energy, and the lifetime of the phonons dissipating \textit{directly} to the gold goes to infinity for higher energy modes. For the modes above the Debye frequency, the dissipation \textit{directly} into the leads approaches zero \cite{pecchia2007, romano2007}.Instead the decay channels into the leads must involve one-to-many processes, such that the phonon energy spreads out on several modes before decaying into the gold. Engulfing ourselves in this kind of calculation, is beyond the scope of this paper, but can in principle be done\cite{nitzan1975}. Instead, $J_q$, is taken as an empirical parameter. From equation \ref{steadystate}, it can be seen that the physical interpretation of this parameter is to bring the system back towards the equilibrium temperature, so that any heating or cooling effect induced from the tunneling current will simply be damped. Previous computational studies on the size of $J_q$, find that this parameter is in the order of $10^8$ to $10^{12}Hz$, for the low energy modes.
\\
In the present study, $J_q = 10^{2}Hz$, to illustrate the balances between the current-induced absorption and emission, we find, however, that the external dissipation is negligible at rates below $10^{10}Hz$
\subsection{Acknowledgement}
The authors thank Olov Karlstr\"{o}m for fruitful discussions. The research leading to these results has received funding from the European Research Council under the European Union's Seventh Framework Program (FP7/2007-2013)/ERC Grant Agreement No. 258806.
\bibliography{refs}
\end{document}